\documentclass[doublecol,figures]{epl2} 

\newcommand{\beq}{\begin{equation}}
\newcommand{\eeq}{\end{equation}}
\newcommand{\ba}{\begin{array}}
\newcommand{\ea}{\end{array}}
\newcommand{\ds}{\displaystyle}
\newcommand{\beqa}{\begin{eqnarray}}
\newcommand{\eeqa}{\end{eqnarray}}
\newcommand{\beqas}{\begin{eqnarray*}}
\newcommand{\eeqas}{\end{eqnarray*}}
\newcommand{\n}{\nonumber}
\newcommand{\eps}{\epsilon}

\newcommand{\f}{\frac}
\newcommand{\ra}{\rightarrow}

\title{On zero energy states in graphene}

\author{C.-L. Ho\inst{1} \and P. Roy\inst{2} }

\institute{                    
  \inst{1} Department of Physics,
Tamkang University,
Tamsui 25137, Taiwan\\
  \inst{2} Physics \& Applied Mathematics Unit,
Indian Statistical Institute,
Kolkata - 700 108, India
}

\abstract{
We obtain zero energy states in graphene for a number of potentials and discuss the relation of the decoupled Schr\"odinger-like equations for the the spinor components with non relativistic $\cal{PT}$ symmetric quantum mechanics.
}

\begin{document}

\maketitle

The dynamics of charge carriers in graphene is governed by $(2+1)$ dimensional massless Dirac equation except that the velocity of light ($c$) is replaced by the Fermi velocity ($v_F$) \cite{novo}. In order for graphene to have practical applications it is necessary to confine or control the motion of the electrons. It is well known that magnetic fields can be used to confine the charge carriers in graphene \cite{magnetic}. On the other hand it is generally believed that because of Klein tunneling electric fields are not useful in confining the charge carriers in graphene. However it has recently been shown that using certain types of electric fields it is indeed possible to confine charge carrier in graphene and zero energy states can be obtained analytically for a number of electric field profiles \cite{zero1}. Also an electric field of the secant hyperbolic type has been found to admit not only zero energy solutions but quasi exact solutions as well \cite{portnoi}. The conditions for existence of zero modes of $(1+1)$ dimensioanal Dirac operator have also been investigated from a mathematical point of view \cite{levitin}. In view of the fact that zero energy states in graphene are of particular interest \cite{zero1}, here we shall employ an alternative method, namely, supersymmetric quantum mechanics \cite{khare,kuru} to analyze the model of Ref. \cite{portnoi} and obtain zero energy states of the secant hyperbolic potential. We shall also obtain zero energy states of two new potentials, one of which is the trigonometric counterpart of the one used in ref \cite{portnoi} while the other is a periodic potential. Another important feature of the present model is that it illustrates how non-relativistic $\cal{PT}$symmetry \cite{pt} appears naturally in relativistic quantum mechanics.

To begin with the motion of electrons in graphene in the presence of a potential is governed by the equation
\beq\label{1}
[v_F(\sigma_xp_x+\sigma_yp_y)]\psi+U(x,y)\psi=E\psi,
\eeq
where $v_F=10^6m/s$ is the Fermi velocity, $\sigma_{x,y}$ are the Pauli spin matrices and $U(x,y)$ is the potential. Here we would consider a potential depending only on the $x$ coordinate and consequently the wavefunction can be taken as
\beq
\psi(x)=e^{ik_yy}\left(\ba{c}\psi_A \\ \psi_B\ea\right).
\eeq
Then from Eq.(\ref{1}) we obtain (in unit of $\hbar=1$)
\beq\label{intert1}
\ds(V(x)-\eps)\psi_A-i\left(\f{d}{dx}+k_y\right)\psi_B=0,
\eeq
\beq\label{intert2}
\ds(V(x)-\eps)\psi_B-i\left(\f{d}{dx}-k_y\right)\psi_A=0,
\eeq
where $V(x)=U(x)/\hbar v_F$ and $\eps=E/\hbar v_F$.

It is interesting to note that Eqs.~(\ref{intert1}) and (\ref{intert2}) are invariant under the following transformations:
\beq
 k_y\to -k_y,~~ \psi_A\leftrightarrow \psi_B.
\eeq
This means that if the spinor $\psi=e^{ik_yy}(\psi_A,\psi_B)^t$ is a solution for $k_y$, then $\psi=e^{-ik_yy}(\psi_B,\psi_A)^t$ is a solution for $-k_y$ (here ``$t$" means transpose).
That is, eignestates with opposite signs of $k_y$ are spin-flipped. 

For $k_y=0$, Eqs.~(\ref{intert1}) and (\ref{intert2}) are invariant under the changes $\psi_A\leftrightarrow \pm \psi_B$, and the respective solutions are
\beqa
\psi(x) &=&\psi_A(x)\left(\ba{c}1\\\pm 1\ea\right),\n
\\
\psi_A(x) &=&\exp\left(\mp i \int^x (V(x)-\eps)\, dx\right).
\eeqa
But in this case, the wave function $\psi(x)$ is normalizable only in a finite domain  if $V(x)$ is real.

\vskip 1cm
 Now let us consider only cases with $k_y\neq 0$. 
Defining $\psi_{1,2}=(\psi_A\pm\psi_B)$ we obtain from Eqs.(\ref{intert1}) and (\ref{intert2})
\beq\label{intert3}
\ds\left(V(x)-\eps-i\f{d}{dx}\right)\psi_1+ik_y\psi_2=0,
\eeq
\beq\label{intert4}
\ds\left(V(x)-\eps+i\f{d}{dx}\right)\psi_2-ik_y\psi_1=0.
\eeq
The equations for the components $\psi_{1,2}$ can be easily obtained from Eqs.(\ref{intert3}) and (\ref{intert4}) and are given by
\beq\label{5}
\left[-\f{d^2}{dx^2}-(V(x)-\eps)^2-i\f{dV(x)}{dx}+k_y^2\right]\psi_1=0,
\eeq
\beq\label{6}
\left[-\f{d^2}{dx^2}-(V(x)-\eps)^2+i\f{dV(x)}{dx}+k_y^2\right]\psi_2=0.
\eeq
Eqs.(\ref{5}) and (\ref{6}) can be interpreted as a pair of Schr\"odinger equations with \emph{energy dependent potentials}
\beq\label{pot3}
V_1(x)=-\left(V(x)-\eps\right)^2-iV^\prime(x)+k_y^2,
\eeq
\beq\label{pot4}
V_2(x)=-\left(V(x)-\eps\right)^2+iV^\prime(x)+k_y^2,
\eeq
where the prime indicates differentiation w.r.t $x$. To relate the above potentials with supersymmetry it may be recalled that a pair of potentials $V_{\mp}(x)$ are called supersymmetric partners if they are of the form \cite{khare}
\beq\label{susy1}
V_-(x,a_1)=W^2(x)-W^\prime(x)+d
\eeq
\beq\label{susy2}
V_+(x,a_1)=W^2(x)+W^\prime(x)+d
\eeq
where $W(x)$ is called the superpotential, $d$ denotes the factorization energy and $a_1$ is a set of parameters. Furthermore supersymmetric partner potentials are translationally shape invariant if
\beq\label{si}
V_+(x,a_1)=V_-(x,a_2(a_1))+R(a_1)
\eeq
where $a_2(a_1)$ and $R(a_1)$ are functions independent of $x$. It turns out that for all the solvable one-dimensional quantal systems (which can be grouped  into ten cases), $a_2$ differs from $a_1$ by only a constant.  Hence these systems are termed translationally shape invariant systems. An important property of shape invariant potentials is that the spectrum can be found in a purely algebraic way \cite{khare}. 
It may be observed that the superpotential can be found from the equation $W^2(x)\pm W^\prime(x)=-V^2(x)\pm iV^\prime(x)$. So comparing Eq.(\ref{susy1}) and Eq.(\ref{susy2}) with Eq.(\ref{pot3}) and Eq.(\ref{pot4}) respectively it is easy to see that the former set of potentials are formally of supersymmetric form if $k_y$ is identified as the factorization energy. 

Now from Eqs.(\ref{pot3}) and (\ref{pot4}) it is seen that $V_{1,2}$ are complex potentials and for any even function $V(x)$ they are in fact $\cal{PT}$ symmetric where \cite{pt}
\beq
{\cal{P}} : x\ra -x,~~~~{\cal{T}} : i\ra -i .
\eeq
So unless $\cal{PT}$ is not spontaneously broken the Schr\"odinger equations (\ref{5}) and (\ref{6}) would admit real eigenvalues. 

We shall now consider some specific examples.  In one-dimensional quantal systems, there exists ten cases of exactly solvable potentials, all of which can be treated by supersymmetric method as discussed in \cite{khare}. But only two cases, when  fit into the form of potentials in Eq.(\ref{pot3}) and Eq.(\ref{pot4}), can provide  energies which are real valued and eigenfunctions which are normalizable.  These we present as Examples 1 and 2 below.  In Example 3, we consider real zero energy band edge solution for a periodic potential.
Physically, the first two potentials provide exactly solvable  potentials which can confine the electrons in graphene, while the third gives an exactly solvable potential in which the electrons 
can move freely in graphene.  These potentials could be fabricated in the laboratory and would be useful to study the transport behavior of electrons in graphene. 

\bigskip

{\bf Example 1.} Here we shall analyze the potential \cite{portnoi}
\beq\label{ex1}
V(x)=-\lambda \,{\rm  sech}~x,~~~~\lambda>0,~~x\in \left(-\infty, \infty\right).
\eeq
This potential has been considered in Ref.\cite{zero1} and quasi exact solutions have also been found \cite{portnoi}. Here we shall consider $\eps=0$ for which the effective potentials become
\beq\label{rm}
\ba{l}
V_1(x)=-\lambda^2  {\rm sech}^2x - i\lambda\ {\rm sech}\, x \tanh\,x +k_y^2\, ,\\
V_2(x)=-\lambda^2 {\rm sech}^2x+ i\lambda\ {\rm sech}\, x \tanh\, x+k_y^2\,.
\ea
\eeq
It is easy to see that the potentials above are $\cal{PT}$ symmetric. These potentials are in fact $\cal{PT}$ symmetrized hyperbolic Scarf II potential of the form \cite{khare}
\beqa\label{RM}
V_{{\rm Scarf II}}(x)&=&-(B^2+A^2+A){\rm sech}^2x\n \\
 && +iB(2A+1){\rm sech}\, x \tanh\,x.
\eeqa
The potential (\ref{RM}) is shape invariant and consequently the associated Schr\"odinger equation, namely,
\begin{equation}
\left(\f{d^2}{dx^2}-V_{{\rm Scarf II}}(x)\right)\,\phi_n (x)=E_n \phi_n(x),
\end{equation}
 is exactly solvable with energy eigenvalues ($E_n$) and wavefunctions ($\phi_n$) given by \cite{khare}
\beq\label{sol1}
\ba{l}
E_n=-(A-n)^2,\\
\phi_n=i^n(1+y^2)^{-A/2}~e^{-iB\tan^{-1}y}\\
~~~~~~\times P_n^{(-B-A-1/2,B-A-1/2)}(iy),\\
y=\sinh\,x,~~~~n=0,1,2....<[A-1],
\ea
\eeq
where $P_n^{(a,b)}(x)$ denotes Jacobi polynomials (hereafter $E_n$ and $\phi_n(x)$ refer to the energies and eigenfunctions of the associated Schr\"odinger equation of a component of the Dirac equation). It is interesting to note that each of the  potentials in (\ref{rm}) is shape invariant although they are not translationally shape invariant partner of one another. Now comparing $V_2(x)$ with the potential in (\ref{RM}) we find
\beq\label{ky}
\ba{l}
\ds A=\lambda-\f{1}{2} >0,~~~~\ds B=-\f{1}{2},\\
k_y^2=-E_n=\left(\lambda-n-\f{1}{2}\right)^2,~~~~n=0,1,2,...< \left[\lambda-\f{3}{2}\right],
\ea
\eeq
and the corresponding solutions are given by
\beq\label{sol}
\ba{l}
\psi_2(x)=i^n(1+y^2)^{-\lambda+1/2}~e^{\f{i}{2}tan^{-1}y} \n \\
~~~~~~~~~~~  \times P_n^{(-\lambda+1/2,-\lambda-1/2)}(iy),
\ea
\eeq
\beq
y=\sinh\,x,~~~~n=0,1,2....< \left[\lambda-\f{3}{2}\right],~~\lambda>\frac12.
\eeq
The wave function $\psi_1$ can be obtained by using Eq.(\ref{sol}) in the intertwining relation Eq.(\ref{intert3}).
Thus we find that the zero energy state is degenerate and the number of degenerate zero energy levels depends on the choice of the parameter $\lambda$. The other interesting point is the fact that the momentum $k_y$ in the $y$ direction can no longer take arbitrary values but is quantized and can only take a certain number of values as dictated by Eq.(\ref{ky}). 

Finally we mention a point about reality of the spectrum. As we started with the real potential (\ref{ex1}) it is obvious that the spectrum would be real. On the other hand the potentials in (\ref{rm}) are $\cal{PT}$ symmetric. It is known
\cite{ahmed} that for the $\cal{PT}$ symmetric potential $V(x)=-v_1 {\rm sech}^2x-iv_2 {\rm sech}\,x\,\tanh\,x$, $\cal{PT}$ symmetry is unbroken if $v_1+\f{1}{4}\geq |v_2|$ while it is spontaneously broken otherwise. In the former case the spectrum is entirely real while in the latter case the eigenvalues occur in complex conjugate pairs. In the present case it is seen that the former condition is always satisfied i.e, $\lambda^2+\f{1}{4} \geq |\lambda|$ and hence the spectrum is always real  as it should be.

\bigskip

{\bf Example 2.}

Here we present the trigonometric counterpart of Example 1. The potential is taken to be
\beq
V(x)=\lambda\, \sec~x,~~~~\lambda>0,~~x\in \left[-\frac{\pi}{2}, \frac{\pi}{2}\right].
\label{V_eg2}
\eeq
In  Fig.~1 we have presented a plot of the potential (\ref{V_eg2}) which clearly shows that it supports bound states.

Then the effective potentials for  $\eps=0$  become
\beq\label{efftrig}
\ba{l}
V_1(x)=-\lambda^2 \sec^2x + i\lambda \sec\,x \tan\,x+k_y^2\, ,\\
V_2(x)=-\lambda^2 \sec^2x - i\lambda \sec\,x \tan\,x+k_y^2\,.
\ea
\eeq
As in Example~1, these potentials too are $\cal{PT}$ symmetric. They are in fact $\cal{PT}$ symmetrized trigonometric Scarf I potential of the form \cite{khare}
\beq\label{Scarf}
V_{{\rm Scarf\,I}}(x)=-(B^2-A^2+A) \sec^2 x- iB(2A-1)\sec\,x \tan x .
\eeq
The potential (\ref{Scarf}) is shape invariant and exactly solvable with energy eigenvalues ($E_n$) and wavefunctions ($\phi_n$) given by \cite{khare}
\beq\label{sol2}
\ba{l}
E_n=(A+n)^2,\\
\phi_n=i(1-y)^{(A-B)/2}(1+y)^{(A+B)/2}~e^{-iB\tan^{-1}y}\\
~~~~~~~~~ \times P_n^{(A-iB-1/2,A+iB-1/2)}(y),\\
y=\sin x,~~~~n=0,1,2,\ldots
\ea
\eeq
Similar to the previous case each of the potentials in (\ref{efftrig})  are shape invariant. Thus comparing $V_1(x)$ with the potential in (\ref{Scarf}) we find
\beq\label{ky2}
\ba{l}
\ds A=\lambda+\f{1}{2},~~~~\ds B=-\f{1}{2},\\
k_y^2=\left(\lambda+n+ \f{1}{2}\right)^2,~~~~n=0,1,2,\ldots,
\ea
\eeq
and the corresponding solutions are given by
\beq
\psi_1(x)\sim \left(\cos\, x\right)^{\lambda-\frac12}  P_n^{(\lambda+i/2,\lambda-i/2)}(\sin\,x),~~~~n=0,1,2.....
\eeq
The zero energy state in this case is infinitely degenerate. The other component of the spinor and hence the original Dirac spinor can be found as in the previous example.

\bigskip

{\bf Example 3.}

Charge carriers under the influence of periodic potentials have been studied by a number of authors to study various properties of graphene \cite{periodic}. For example, a superlattice potential of the form $V(x)=V_0\cos(G_0x)$ has been used to study transport properties in graphene \cite{brey1}. Here we consider a periodic potential of the type
\beq\label{period}
V(x)=-\lambda \sin(2x),~~~~  -\infty < x <\infty.
\label{V_eg3}
\eeq
In Fig.~2 we have presented a plot of the potential (\ref{period}) and the cosine potential of ref \cite{brey1}.

  In the present case the effective potentials are given by
\beq
V_{1,2}(x)=-\lambda^2 \sin^2(2x) \pm 2i\lambda \cos(2x)+k_y^2.
\eeq
The above potentials are $\cal{PT}${\footnote[1]{Here parity transformation is defined as $x\ra x+\pi/2$} invariant periodic potentials of the form
\beq\label{t3}
U(x)=-b^2 \sin^2(2x)+2iab~ \cos(2x).
\eeq
It is known that for the potential (\ref{t3}) $\cal{PT}$ symmetry is broken when $a$ is an even integer while it is unbroken if $a$ is an odd integer \cite{khare1}. Furthermore when $a$ is an odd integer there are $a$ quasi exactly solvable states of period $\pi$ and there are at most $(a+1)/2$ band gaps of period $\pi$ \cite{khare1}. Now comparing $V_1(x)$ with (\ref{t3}) we find that in the present example $b=\lambda, a=1$. The zero energy solution is given by \cite{khare2}
\beq
\psi_1(x)=e^{-i\lambda \cos(2x)},~~~~k_y=\pm\sqrt{1-\lambda^2}.
\eeq
Thus we find that for the solution to be valid $\lambda<1$ and also $k_y$ can only assume specific values. 

In summary,  we have obtained exact analytical zero energy solutions of number of potentials some of which have not been considered before within the context of Dirac equation. It has also been shown how the model can be related to a non relativistic supersymmetric quantum mechanical system. In this context we would like to mention that there are many other potentials which may support bound states but for which equations of the form (\ref{5}) or (\ref{6}) may not admit exact solutions. In such cases it may be possible some techniques like the SWKB method to obtain the energy eigenvalues. Another interesting aspect of the model is its relation to $\cal{PT}$ symmetry. So far $\cal{PT}$ symmetry has been realized in the field of optics. However the present model provides a scenario where $\cal{PT}$ symmetry appears naturally.
\acknowledgments

The work is supported in part by the Ministry of Science and Technology (MOST)
of the Republic of China under Grant NSC-102-2112-M-032-003-MY3 (CLH). 
PR wishes to
thank the R.O.C.'s National Center for Theoretical Sciences (North Branch)  and
National Taiwan University for supporting a visit during which part of the work was done.



\vskip 2cm

\centerline{\bf Figure caption}
\medskip
\begin{enumerate}
\item[]
Fig.~1:Plot of the potential (\ref{V_eg2}) for  $\lambda=1$.

\item[]
Fig.~2:
Plot of the potential $V(x)=V_0 cos(G_0 x)$
of \cite{brey1} 
for
$V_0=1, G_0=2$ (dotted curve) and potential in (\ref{V_eg3}) 
for $\lambda=1$ (solid curve).
\end{enumerate}

\end{document}